# *Multiresonance and chaotic behavior analysis for polarization in material modeled by multifrequency excitations duffing oscillator*

**C. AINAMON[†], C. H. MIWADINOU[ǂ], V.A. MONWANOU[ǂ] & J. B. CHABI OROU[ǂ,‡]**

[†] Ecole Doctorale Sciences des Matériaux, Université d'Abomey – Calavi, Bénin

[ǂ] Institut de Mathématiques et de Sciences Physiques, Université d'Abomey – Calavi, Bénin.

[‡] Département de Physique, Université d'Abomey – Calavi, Bénin

Correspondance : J. B. CHABI OROU, Institut de Mathématiques et de Sciences Physiques, 01 BP 613 Porto- Novo, Bénin.

E-mail : jchabi@yahoo.fr ou jean.chabi@imsp-uac.org

**Abstract:**

This paper considers nonlinear dynamics of polarization oscillations when some materials when they are subjected to the action of an electromagnetic wave modeled by multifrequency forced Duffing equation. Multiresonance and chaotic behavior are analysed. For analysis of the case of resonance, the method of multiple scales is used and it has been found from the equation of the amplitudes for each of the possible resonance system. Possible resonances are inter alia the resonances or sub superharmonic, the primary resonance and other resonances called secondary. The phenomena of amplitude jump and hysteresis for polarization were observed and analyzed. Finally, the study of chaotic behavior for polarization was made by numerical simulation using the Runge-Kutta fourth order.

## 1- Introduction

We consider the Duffing oscillator equation which modelize the material submitize at multfrequency excitation electromagnets wave [1]

$$\ddot{P} + 2\varepsilon\mu\dot{P} + P + \varepsilon P^3 = F_1 \cos\omega_1 t + F_2 \cos\omega_2 t \qquad (1)$$

The interaction of light with a non-linear optical material changes the properties of the material and thereby alters the frequency, phase or polarization of the light passed through. This is the cause of significant non-linear optical phenomena including the optical Kerr effect and frequency doubling [5]. Resonance plays an important role in non-linear physics problems specifically in the study of non-linear behavior of a material. The response of a nonlinear system to a weak periodic signal can be enhanced by means of an appropriate noise [14], a high frequency periodic force [15], or a chaotic signal [16]. The enhancements of the response of a system due to the applied weak noise, a high frequency force, or a chaotic signal are termed as stochastic resonance, vibrational resonance, and chaotic resonance, respectively. Among them, much attention has been given to stochastic resonance and a lot of progress has been made. However, the analysis of multiresonance has also received considerable interest in recent years due to its importance in a wide variety of contexts in physics, engineering, and biology. On the other hand, so far, it has been seen that the presence of chaos in many systems has been extensively demonstrated and is very common. The purpose of this paper is to take into account the multiresonance of in the modeling of (1) and then to investigate using analytical methods, harmonic and resonant states which can be displayed by the model in such conditions . We also aim to perform the possible bifurcation mechanisms of the model using numerical tools. The phase portraits and basin of attraction corresponding to the bifurcation diagram and its corresponding Lyapunov.

The paper is organized as follows: In Sec. 2, addresses the multiresonant states of the model through the multiple time scales method [4]. In Sec. 3, we will point out, bifurcation and chaotic sequences of the model. We conclude in Sec. 4.

2-**Multiresonant states**

Generally, may type of oscillations can be found in a forced system additionally, to the harmonic oscillatory states. Such oscillation occurs when the external frequency is too close or far from the internal frequency, according also to the external excitation strength. Since these oscillations rise up at different time scale, the best tool to be use for their investigation is the multiple time method [4]. In such a situation, an approximate solution is generally sought as follows

$P(t,\varepsilon) = P_o(T_o, T_1) + \varepsilon P_1(T_o, T_1) + \cdots$ (2)

where the fast time scale to end the slow time scale are associated respectively to the unperturbed system and to the amplitude and phase induced by the global first order perturbation. The first and second time derivatives can now be rewritten as follow :

$\frac{d}{dt} = = D_o + \varepsilon D_1 + \ldots, \frac{d^2}{dt^2} = D_o^2 + 2\varepsilon D_o D_1 + \varepsilon^2 D_1^2 + \cdots$ (3)

With $D_n = \frac{\partial}{\partial T^n}$, $T_n = \varepsilon^n t$; $n = 0, 1, 2, \ldots \ldots$

## 2-1 Primary resonant state

In this state, we put that $F_1 = \varepsilon F_1, F_2 = \varepsilon F_2$
the external frequencies is given by $\omega_1 \simeq \omega_2 \simeq 1 + \varepsilon\sigma$
Inserting these relations in Eq. (4), we obtained
$$D_o^2 P_o + P_o = 0 \quad (4)$$
and
$$D_o^2 P_1 + \omega_0 P_1 = \left[-2 i \omega_0 A' - 2 i\mu \omega_0 A - 3\lambda A^2 \bar{A} + \frac{1}{2} F_1 e^{i\sigma T_1} + \frac{1}{2} F_2 e^{i\sigma t_1}\right] e^{i\omega_0 T_o} + NST + CC \quad (5)$$

Equating a secular term at 0, we obtain:
Where $F = F_1 + F_2$, $\omega_0 = 1$. Soit
$$-2i\ A' - 2 i \mu A - 3\lambda A^2 \bar{A} + \frac{F}{2} e^{i\sigma T_1} = 0 \quad (6)$$
Injecting, A in (6) separating real and imaginary terms, we obtain the following system equation:

$$a' = -\mu a + \frac{F}{2} \sin\emptyset \quad (7)$$
$$and \quad a\theta' = \frac{3}{8} \lambda a^3 - \frac{F}{2} \cos\emptyset \quad (8)$$

Where $\emptyset = \sigma T_1 - \theta$

Finally, we obtain the coupled flow for the amplitude and phase
$$a' = -\mu a + \frac{F}{2} \sin\emptyset \quad (9)$$

$$a\emptyset' = a\sigma - \frac{3}{8} \lambda a^3 + \frac{F}{2} \cos\emptyset \quad (10)$$

The steady state conditions ( $a' = \emptyset' = 0$ )

These conditions implies:
$$\mu a = \frac{F}{2} \sin\emptyset \quad (11)$$
$$a\sigma - \frac{3}{8} \lambda a^3 = -\frac{F}{2} \cos\emptyset \quad (12)$$

Eliminating the phase $\emptyset$ in Eq (11) and (12), we obtain the following nonlinear algebraic equation

$$\mu^2 a_0^2 + \left(a_0 \sigma - \frac{3}{8} \lambda\ a_0^3\right)^2 = \frac{F^2}{4} \qquad (13)$$

where $a_0$ and $\emptyset_0$ are respectively the values of $a$ and $\emptyset$ on the steady-state is of interest only if it is stable.

Figure 1 shows a representative curve, called a frequency-response curve, for the variation of a with $\sigma$. The bending of the frequency-response curve is responsible for a jump phenomenon. To explain this, we imagine that an experiment is performed in which the amplitude of the excitation is held fixed, the frequency of the excitation (i.e., $\sigma$) is very slowly varied up and down through the linear natural frequency, and the amplitude of the harmonic response is observed. The experiment is started at a frequency corresponding to point A on the curve in Figure 1. As frequency is reduced, $\sigma$ decreases and a slowly increases through point B until point C is reached. As $\sigma$ is decreased further, a jump from point C to point D takes place with an accompanying increase in a, after which a decreases slowly with decreasing $\sigma$. If the experiment is started at point E and $\sigma$ is increased, a increases slowly through point D until point F is reached. As $\sigma$ increased further, a jump from point F to B takes place with an accompanying decrease in a, after which a decreases slowly with increasing $\sigma$. The maximum amplitude corresponding to F is attainable only when approached from a lower frequency. The portion of the response curve between points C and F is unstable, and hence, cannot be produced experimentally.

If the experiment is performed with the frequency of the excitation $\omega_1 = \omega_2$ held fixed while the amplitude of excitation is varied slowly, a similar jump phenomenon can be observed. Suppose that the experiment is started at point A in Figure 2. As $F_1+F_2$ is increased, a slowly increases through point B to point C. As $F_1+F_2$ is increased further, a jump takes place from point C to point D, with accompanying increase in a, after which a increases slowly with $F_1+F_2$. If the process is reserved, a decreases slowly as $F_1+F_2$ decreases from point E to point F. As $F_1+F_2$ is decreased further, a jump from point F to point B takes place, with an accompanying decrease in a, after which a decreases slowly with decreasing $F_1+F_2$. Figure1 (b) and Figure2 (b) show the effect on jump and hysteresis phenomena of F= $F_1+F_2$ and $\mu$ respectively. Through these figure we notice that the excitation amplitude or damping parameter can used to controlled the polarization oscillation in material.

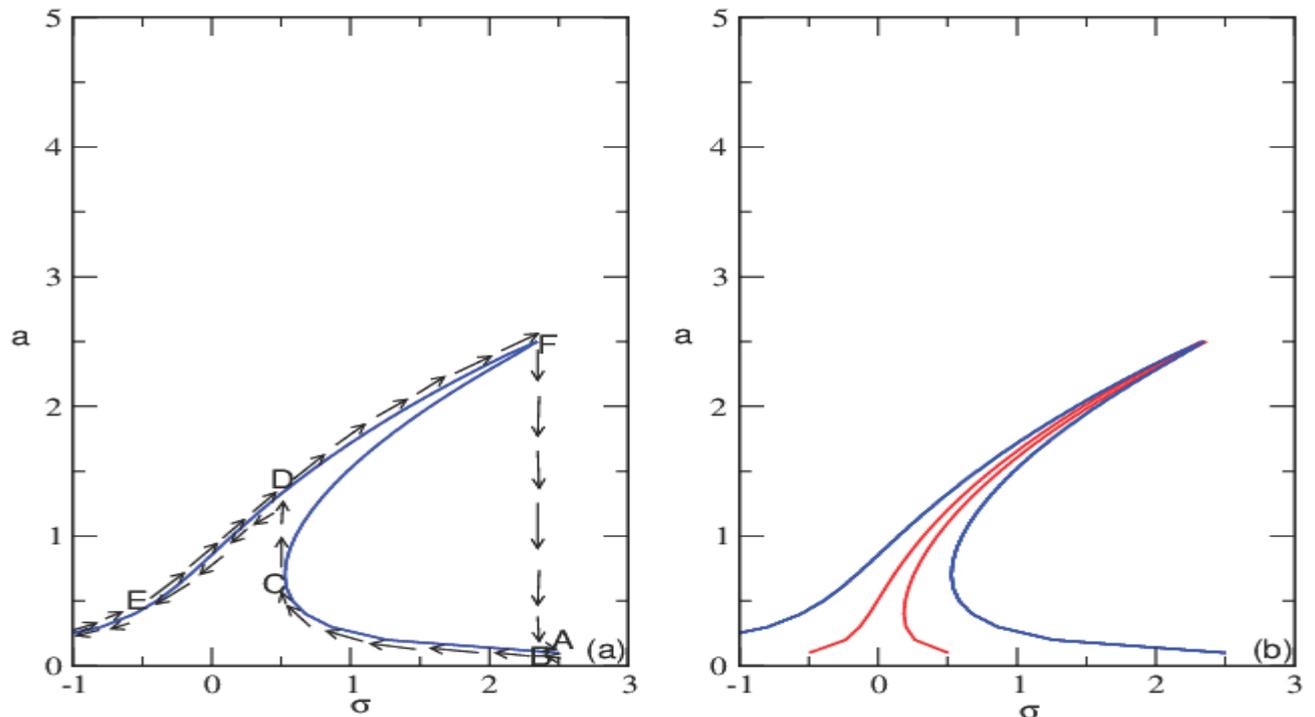

**Figure1**: Jump phenomena for primary resonance of the polarization in the system for $\mu = 0.004, \; \lambda = 1;\, with\, (a): F_1 = F_2 = F = 0.14$ and (b): The effects of parameter F.

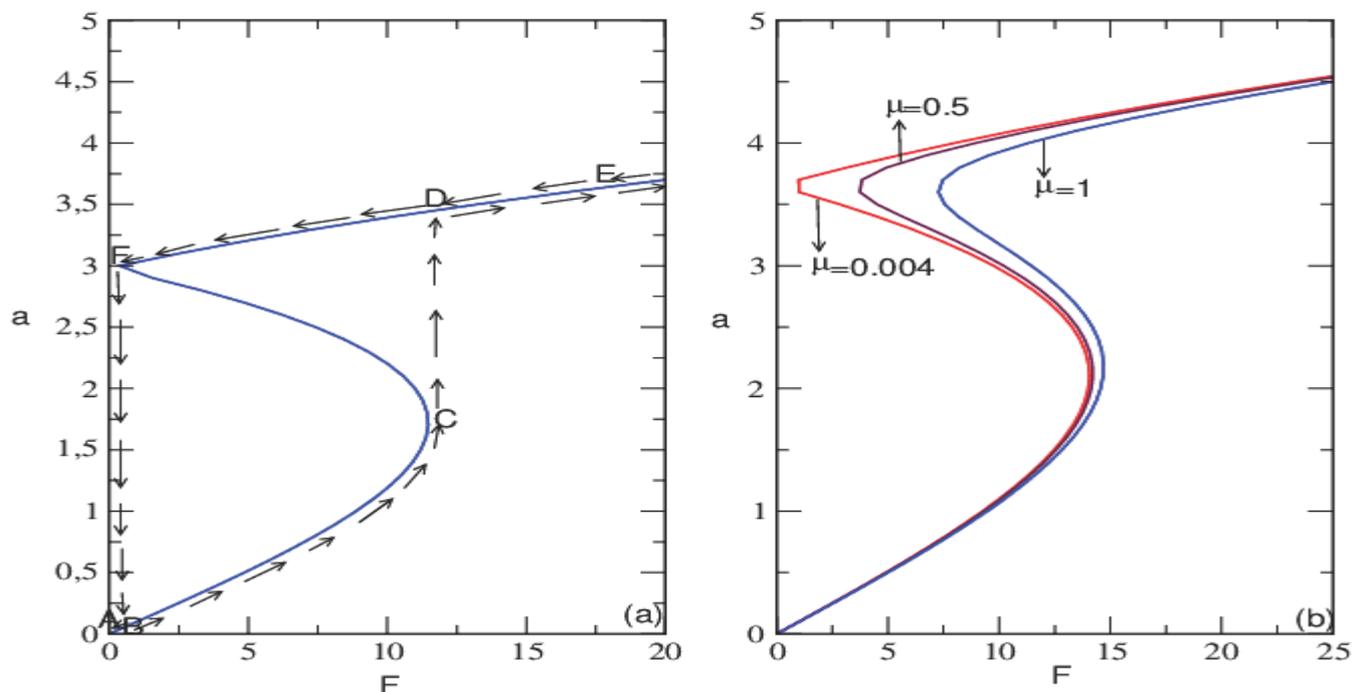

**Figure2**: (a):Jump phenomena for primary resonance of the polarization in the system for the parameters of figure1 and (b): The effects of parameter $\mu$.

### 2-2 Other resonant states

When the amplitude of the sinusoidal external force is lager, other type of oscillations can be displayed by the model, namely the superharmonic, the subharmonic oscillatory states [1] and other secondary resonances states. It is now assumed that $F = \varepsilon^0 F$ and therefore, one obtains the following equations at different order of $\varepsilon$

Order $\varepsilon^0$
$$D_0^2 P_0 + P_0 = F_1 \cos \omega_1 t_o + F_2 \cos \omega_2 t_o \quad (14)$$
Oder $\varepsilon^1$
$$D_0^2 P_1 + P_1 = -2 D_0 D_1 P_0 - 2\mu D_0 P_0 - \lambda P_0^3 \quad (15)$$

From Eq(14), we have

$$P_0 =$$
$$+ A e^{i T_0} + \bar{A} e^{-i T_0} + \Lambda_1 e^{i\omega_1 T_0} -$$
$$3 \lambda \bar{A} \Lambda_1^2 e^{i\omega_1 T_0} + \Lambda_2 e^{i\omega_2 T_0} \quad \Lambda_2 e^{-i\omega_2 T_0} \quad (16)$$

With

$$\Lambda_1 = \frac{F_1}{2(1-\omega_1^2)}, \quad \Lambda_2 = \frac{F_2}{2(1-\omega_2^2)}, \quad \text{and} \quad A = \frac{1}{2} a(T_1) e^{i\theta} \quad (17)$$

With substituting the general solution $P_0$ into Eq. (15) yields[1]

$$D_0^2 P_1 + P_1 = [-2i (A' + \mu A) - 3\lambda A^2 \bar{A} - 6\lambda \Lambda_2^2 A - 6\lambda A \Lambda_1^2 ] e^{iT_0}$$
$$+ [-2i \mu \Lambda_1 - 6\lambda A \bar{A} \Lambda_1 - 6\lambda \Lambda_1 \Lambda_2^2 - 3\lambda \Lambda_1^3] e^{i\omega_1 T_0}$$
$$+ [-2i \mu \Lambda_2 - 6\lambda \Lambda_2 A \bar{A} - 3\lambda \Lambda_2^3 - 6\lambda \Lambda_2 \Lambda_1^2] e^{i\omega_2 T_0}$$
$$+ - 3\lambda \Lambda_2 \bar{A}^2 e^{i(-2+\omega_2)T_0} - 3\lambda \bar{A} \Lambda_2^2 e^{i(-1+2\omega_2)T_0}$$
$$+ - 3\lambda \Lambda_1 \bar{A}^2 e^{i(-2+\omega_1)T_0} - 3\lambda A^3 e^{i3 T_0} - 3\lambda \Lambda_1^3 e^{i3 \omega_1 T_0}$$
$$- 3\lambda \Lambda_2^3 e^{i3 \omega_2 T_0} - 6\lambda \Lambda_1 \Lambda_2 e^{i(1-\omega_1+\omega_2)T_0}$$
$$- 6\lambda A \Lambda_1 \Lambda_2 e^{i(1+\omega_1-\omega_2)T_0} - 6\lambda \bar{A} \Lambda_1 \Lambda_2 e^{i(-1+\omega_1+\omega_2)T_0}$$
$$- 3\lambda \Lambda_2 A^2 e^{i(2-\omega_2)T_0} - 3\lambda \Lambda_1 A^2 e^{i(2-\omega_1)T_0}$$
$$+ - 3\lambda \Lambda_1 \Lambda_2 e^{i(2\omega_2-\omega_1)T_0} - 3\lambda \bar{A} \Lambda_1^2 e^{i(-1+2\omega_1)T_0}$$
$$- 3\lambda \Lambda_2 \Lambda_1^2 e^{i(-\omega_2+2\omega_1)T_0} - 3\lambda \bar{A} \Lambda_2^2 e^{i(-1+2\omega_2)T_0} + CC + NST \quad (18)$$

where CC denotes the complexe conjugated and NST the non-secular terms.

For the particular solutions of $P_1$ contain secular and small divisions terms. For the uniform expansion, the secular and small division terms must be eliminates by proper chooses depend on type of resonances present. The cases of superharmonic ($3\omega_1 = 3\omega_2 = 1 + \sigma\varepsilon$) and *and suharmonic* ( $\omega_1 = \omega_2 = 3 + \sigma\varepsilon$ are treated in previous work [13]. In this work, we treat the three cases for secondary resonances: $\omega_1+\omega_2 = 2$, $2\omega_2 - \omega_1 = 1$ *and* $2\omega_1 - \omega_2 = 1$

2-2-1 **The case** $\omega_1+\omega_2 = 2$

In this case, we assume that $\omega_1+\omega_2 = 2$ and no other resonances exist to first order. To describe quantitatively the nearness of $\omega_1+\omega_2 = 2$, we introduce the detuning parameter $\sigma$ defined by $\omega_1+\omega_2 = 2+ \sigma\varepsilon$

In this case, Eq.(18) can be rewritten as follow

$$D_0^2 P_1 + P_1 = (-2i(A' + \mu A) - 3\lambda A^2 \bar{A} - 6\lambda \Lambda_2^2 - 6\lambda \Lambda_1^2 A) e^{iT_0} - 6\lambda \bar{A} \Lambda_1 + \Lambda_2 e^{i(-1+\omega_1+\omega_2)T_0} + CC + NST \quad (19)$$

With this condition, we rewrite Eq.(23) as

$$D_0^2 P_1 + P_1 = (-2i(A' + \mu A) - 3\lambda A^2 \bar{A} - 6\lambda A \Lambda_2^2 - 6\lambda \Lambda_1^2 A) e^{iT_0} - 6\lambda \bar{A} \Lambda_1 \Lambda_2 e^{i\sigma T_1} e^{iT_0} \quad (20)$$

Where NST stands for terms do not produce secular terms in $P_1$

Eliminating the secular terms from equation (25) demand that
$$(-2i(A' + \mu A) - 3\lambda A^2 \bar{A} - 6\lambda \Lambda_2^2 - 6\lambda \Lambda_1^2 A e^{iT_0} - 6\lambda \bar{A} \Lambda_1 + \Lambda_2 e^{i\sigma T_1} = 0 \quad (21)$$

Equation Real and imaginary parts to 0 we obtain:
$$\begin{cases} -a' + -\mu a - 3\lambda \ \Lambda_1 \Lambda_2 \ a\cos\emptyset = 0 \\ a\theta' - \frac{3}{8}\lambda a^3 - 3\lambda(\Lambda_1^2 + \Lambda_2^2)a - 3\lambda \ \Lambda_1 \Lambda_2 \sin\emptyset = 0 \end{cases} \quad (22)$$

with $\emptyset = \sigma T_1 - 2\theta_1$

The system Equation (22), become:
$$a' = -\mu a - 3\lambda \ \Lambda_1 \Lambda_2 \ a\cos\emptyset$$

$$a\theta' = a\sigma - \frac{6}{8}\lambda a^3 - 6\lambda(\Lambda_1^2 + \Lambda_2^2)a - 6\lambda \ \Lambda_1 \Lambda_2 \sin\emptyset \quad (23)$$

$a' = 0$ *and* $\emptyset' = 0$ implies

$$- 3\mu\ a_s - 6\lambda\ \Lambda_1\ \Lambda_2\ a_s\ cos\emptyset\ = 0$$

$$a_s\ \sigma\ -\ \frac{6}{8}\lambda\ a_s^3 - 6\lambda(\Lambda_1^2 + \Lambda_2^2)\ a_s = -6\lambda\ \Lambda_1\ \Lambda_2\ a_s\ sin\emptyset \quad (24)$$

The amplitude of this resource state is governed by the following nonlinear algebraic equation

$$4\mu^2 a_s^2\ +\ (a_s\sigma - \frac{6}{8}\lambda\ a_s^3 - 6\lambda(\Lambda_1^2 + \Lambda_2^2)\ a_s)^2 = 36\lambda^2\ \Lambda_1^2\ \Lambda_2^2\ a_s^2 \quad (25)$$

Equation(26) can be rewritten as follow

$$4\mu^2 + (\sigma - \frac{3}{4}\lambda\ a_s^2 - 6\lambda(\Lambda_1^2 + \Lambda_2^2)\ a_s)^2 = 36\lambda^2\ \Lambda_1^2\ \Lambda_2^2 \quad (27)$$

$$\emptyset_s = \frac{1}{2}6\ ET - \frac{1}{2}\tan^{-1}\{-\frac{1}{2\mu}[6 - \frac{3}{4}\lambda\ a_s^2 - 6\lambda(\Lambda_1^2 + \Lambda_2^2)\ ]\} \quad (28)$$

Then $A_s = \frac{1}{2}\ a_s \exp i\ \theta_s$

*Where $a_s$ and aand $\theta_s$ are defined respecty by (27) and (28)*

We conclude that in the order 1, the polarization is defined by
$$P_0 = a_s\ \exp i\ \theta_s\ \cos(t) + \frac{F_1}{1-w_1^2}\cos w_1 t + \frac{F_2}{1-w_1^2}\cos w_2 t \quad (29)$$
Where $a_s$ and $\theta_s$ defined respectively by (27) and (28)
For this order, the susceptibity is defined through the relation
$$P \simeq E_0\chi\ (w_i)E(\ w_i), i = 1,2 \quad (30)$$
$\chi\ (w_i) = \frac{P}{E_0\ E(w_i)}$ Where P is defined by (31)

In Figure 3 we plotted the amplitude-response curve. It is found that with this resonant state, the polarization of material present the highly hysteresis phenomenon. These can induced dangerous or undesirable behaviors in material because the polarization and susceptibity of material presented the singularity which provoked chaotic or catastrophic behaviors.

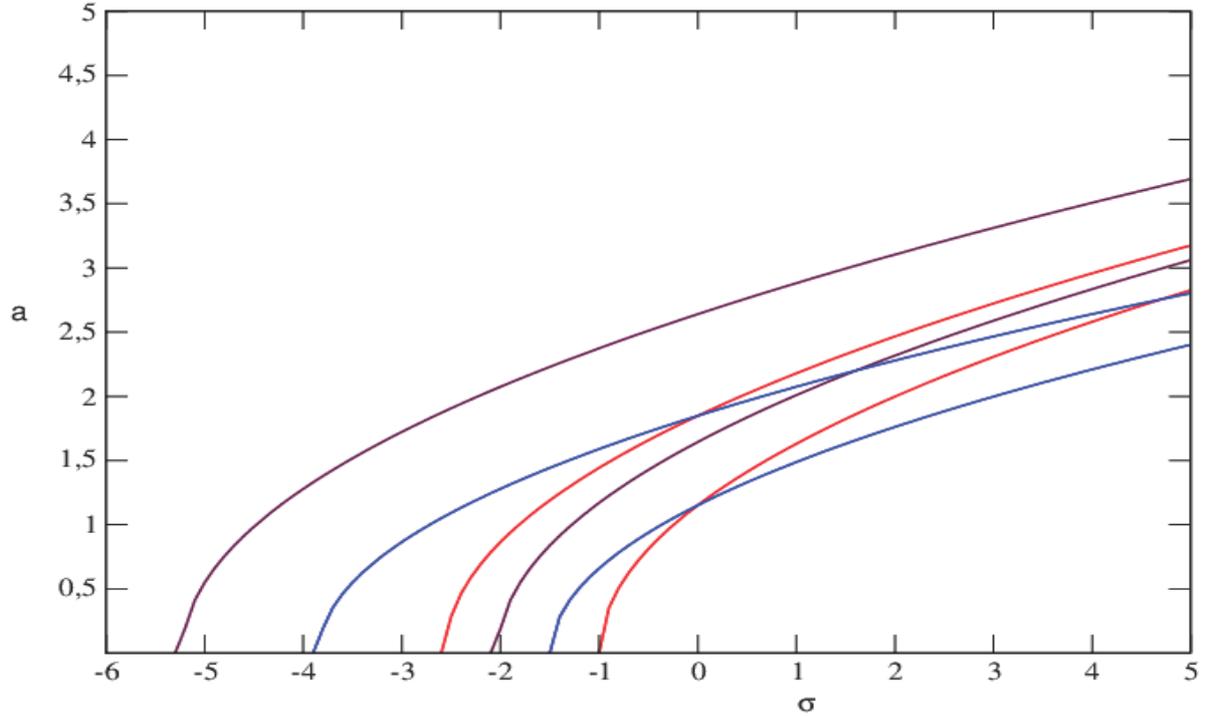

**Figure3**: Effects of amplitude of external forced on the resonance case $\omega_1 + \omega_2 = 2$.

**2-2-2 The cases $2\omega_1 - \omega_2 = 1 + \varepsilon\sigma$ or $2w_2 - w_1 \simeq 1$**

In this case, we assume that $2w_2 - w_1 \simeq 1 + \sigma\varepsilon$

Thing this relation, Eq. (21) can be rewritten as follow:

$D_o^2 P_1 + P_1 = (-2i(A' + \mu A) - 3\lambda A^2 \bar{A} - 6\lambda \wedge_2^2 A - 6\lambda \wedge_1^2 A) e^{iT_0} - 3\lambda \wedge_1 \wedge_2^2 e^{i\sigma\varepsilon T_0} \times e^{iT_0} + cc + NST$ (32)

Eliminating the secular terms from Eq.(32) and
Equating real and imaginary parts to 0, we obtain

$$a' = -\mu a - 3\lambda A^2 \bar{A} \cos\emptyset$$
$$\text{and}$$

$a\theta' = \frac{3}{8} \lambda a^3 + 3\lambda(\wedge_1^2 + \wedge_2^2)a + 3\lambda \wedge_1 \wedge_2^2 \sin\emptyset$ (33)

With

$\emptyset = \sigma T_1 - \theta$

Steady- state solution are seecked, the amplitude of this resonance state is governed by the nonlinear algebraic equation

$\mu^2 a_s^2 + [\sigma - \frac{3}{8}\lambda a_s^2 - 3\lambda(\wedge_1^2 + \wedge_2^2)]^2 - 9\lambda^2 \wedge_1^2 \wedge_2^4 = 0$ (34)

In Figure 4 we plotted the amplitude-response curve. We found the same behaviors as the precedent case.

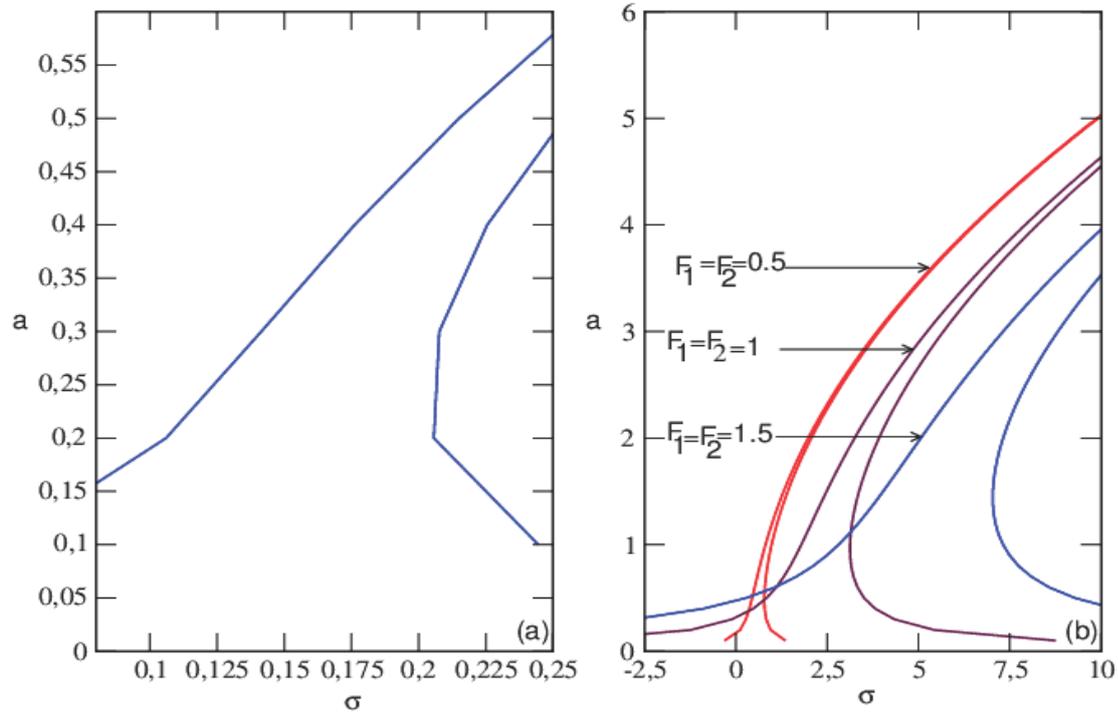

**Figure**:4 Jump phenomena for resonance states $2 w_2 - w_1 \simeq 1$ or $2 w_1 - w_2 \simeq 1$ of the polarization in the system for the parameters of figure1; (a): $w_2 = 1.5$; $w_1 = 2$ and (b): The effects of parameter F

### 3-Bifurcation and chaotic behaviors

To illustrate the dynamical behaviour of the system it is necessary to simulate the original equations. Here we have used the Runge-Kutta method of the forth order to solve Eq.(1) for resonant states to identify the chaotic motion. Draw the resulting bifurcation diagram and the variation of the corresponding largest Lyapunov exponent as the amplitude $F=F_1=F_2$, the parameters of system are varied. The Lyapunov exponent is defined as

$$Lya = \lim_{t \to \infty} \frac{\ln \sqrt{dP^2 + dv_P^2}}{t} \qquad (35)$$

where $dP$ and $dv_P$ are the variations of $P$ and $\dot{P}$ respectively.

For the set of parameters $\mu = 0.004$, $\lambda = 1$, $\varepsilon = 0.271$ and $\omega_1 = \omega_2 = 1$, $\omega_1 + \omega_2 = 2$ and $2\omega_1 - \omega_2 = 1$. In our numerical code we started calculations from the same initial conditions ($P_0 = 1, \dot{P}_0 = 1$). It is found that the model can switch from quasi-periodic oscillations to chaotic states or from chaotic to quasi-periodic oscillations when the amplitude of external excitation is varied as shown in the bifurcation diagram and its corresponding Lyapunov (see Figures 5, 6 and 7). In order to illustrate such situations, we have represented the phase portraits using the parameters of the bifurcation diagrams. With appropriate choice of the amplitude of external excitation F, the phase portraits (chaotic and quasi-periodic motion) are observed in Figures 8, 9 and 10 respectively for primary resonant state (Figure 5), other resonant states $\omega_1 + \omega_2 = 2$ (Figure 6) and $2\omega_1 - \omega_2 = 1$ (Figure 7). We noticed that the polarization model is very chaotic on the large domain in the last case (see Figures 7 and 10). Among these domains, we have the following: $F \in [1.02, 4.46] \cup [5.31, 11] \cup [13, 30]$. It should be emphasized from Figure 7 that there are some domains where the Lyapunov exponent does not match very well the regime of oscillations expected from the bifurcation diagram. Far from being an error which has occurred from the numerical simulation process, such a behavior corresponds to what is called the intermittency phenomenon. Since the model is highly sensitive to the initial conditions, it can leave quasi-periodic state for a chaotic state without changing the physical parameters. Therefore, its basin of attraction has been plotted (see Figure13) in order to situate some regions of initial conditions for which chaotic oscillations are observed. In that figure, the blue or black zone stands for the area where the choice of the initial conditions lead to a chaotic motion while the white area is the domain of periodic or quasi-periodic oscillations. The same analysis are noticed through Figures 11 and 12 while Figure 14 show basin of attraction of polarization oscillations for subharmonic. From these chaotic basin, we notice that with these resonant states, the polarization of material present dangerous or undesirable behaviors.

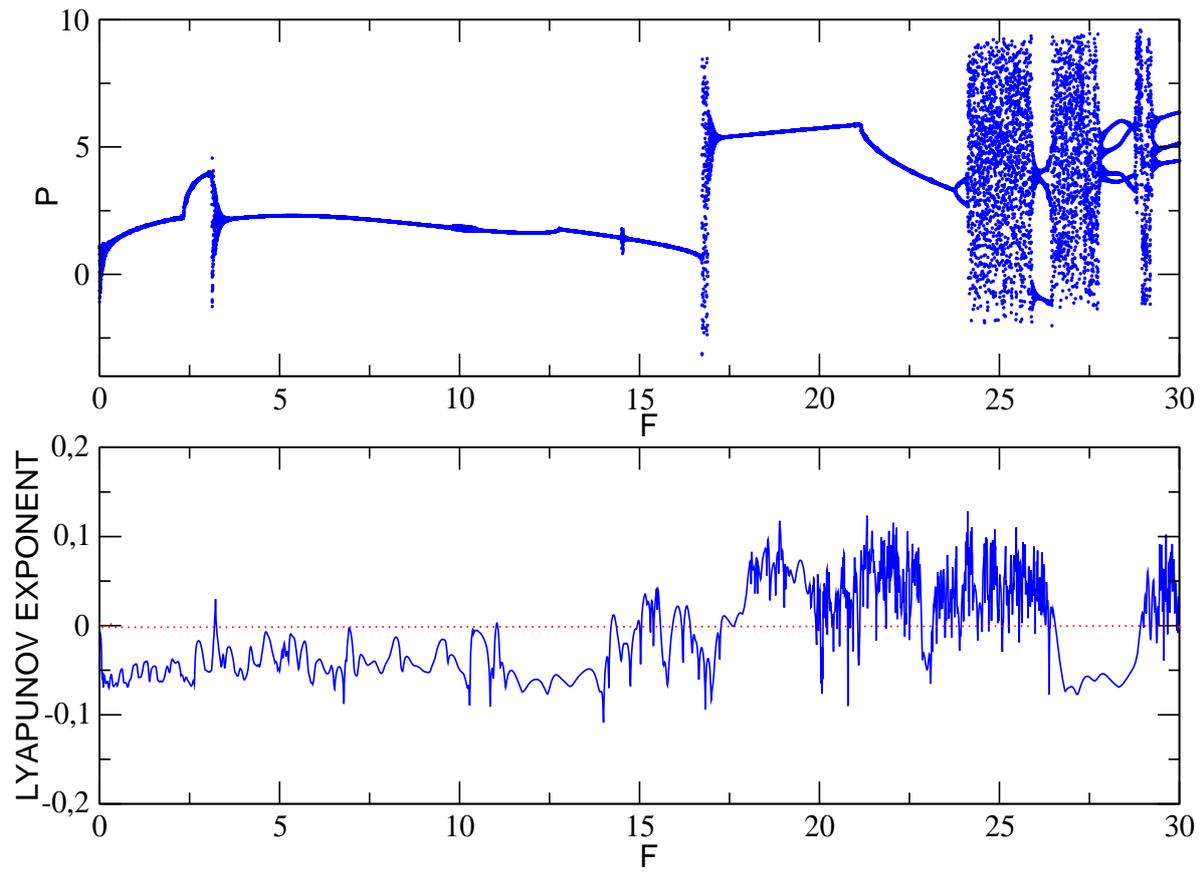

$0.271$ and $\omega_1 \simeq \omega_2 \simeq 1$.

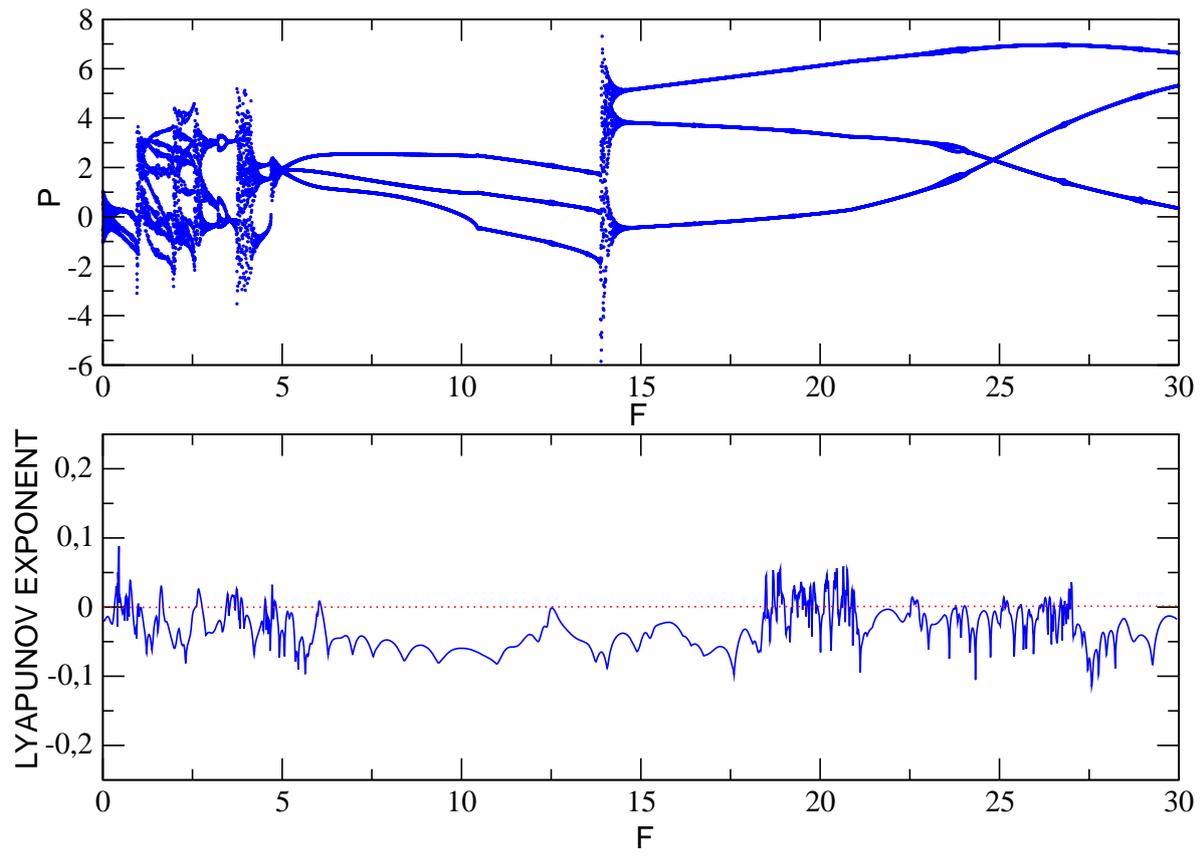

0.271 and $\omega_1 + \omega_2 \simeq 2$.

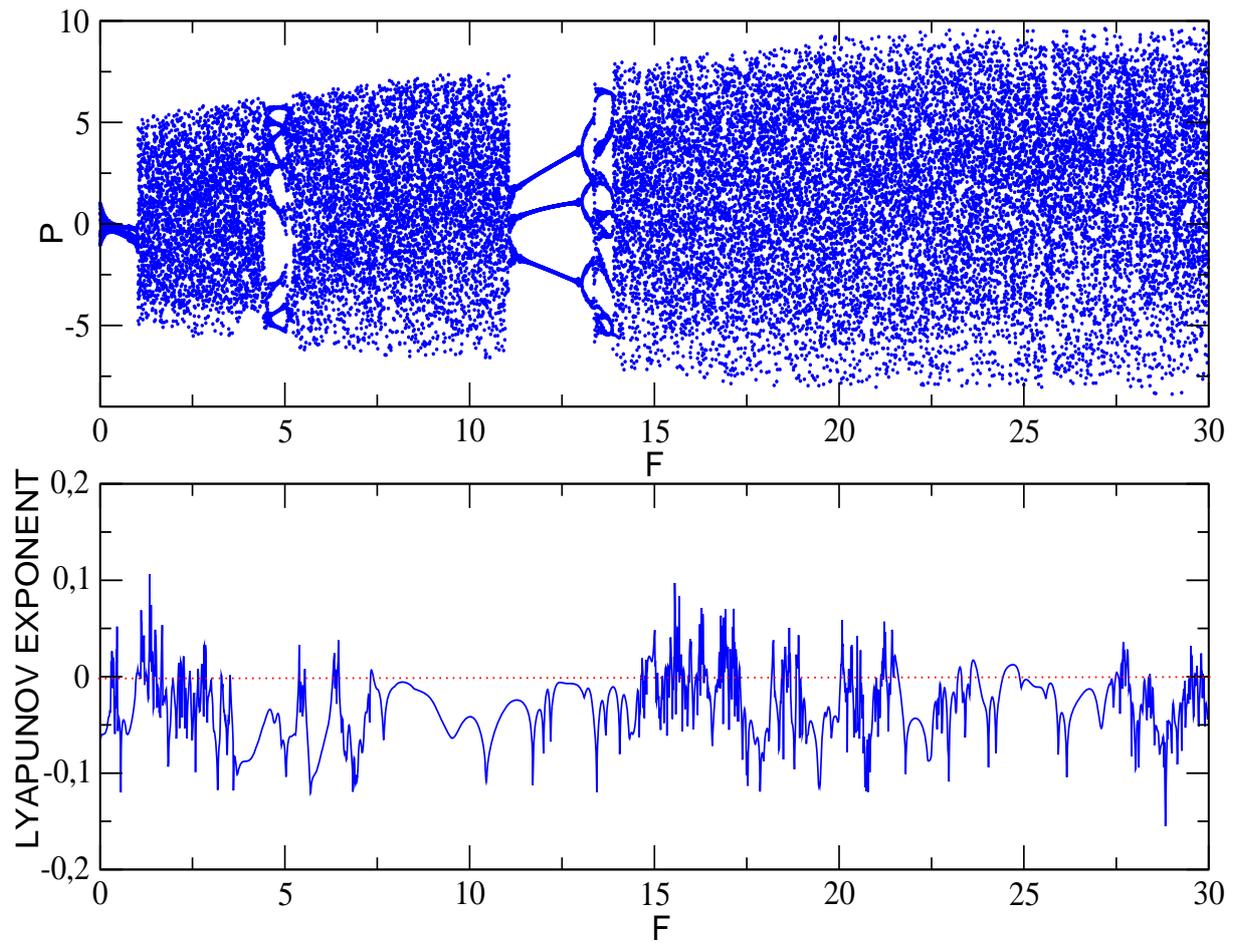

$0.271$ and $2\omega_1 - \omega_2 \simeq 1$.

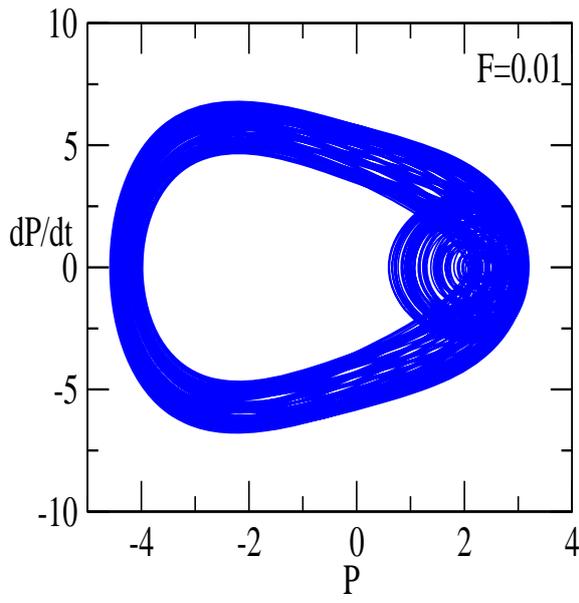 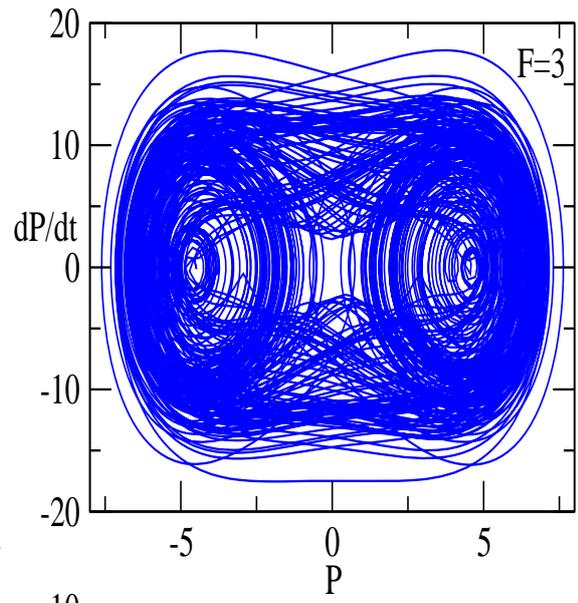
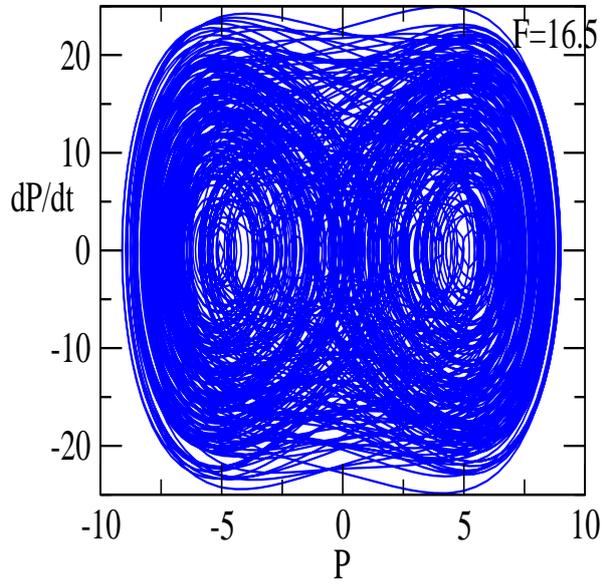 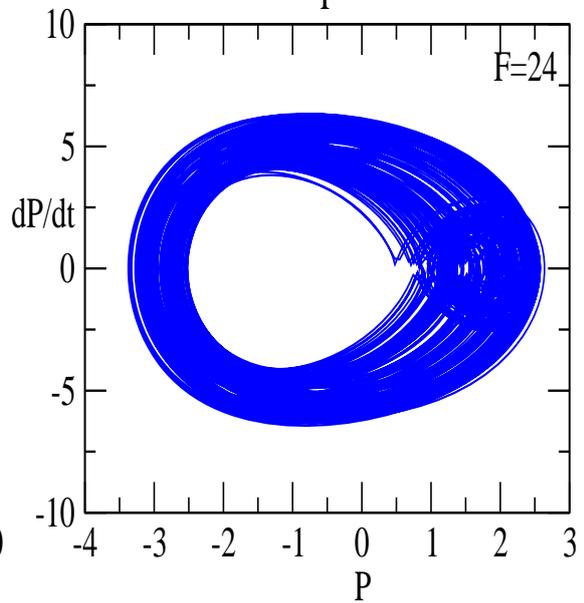

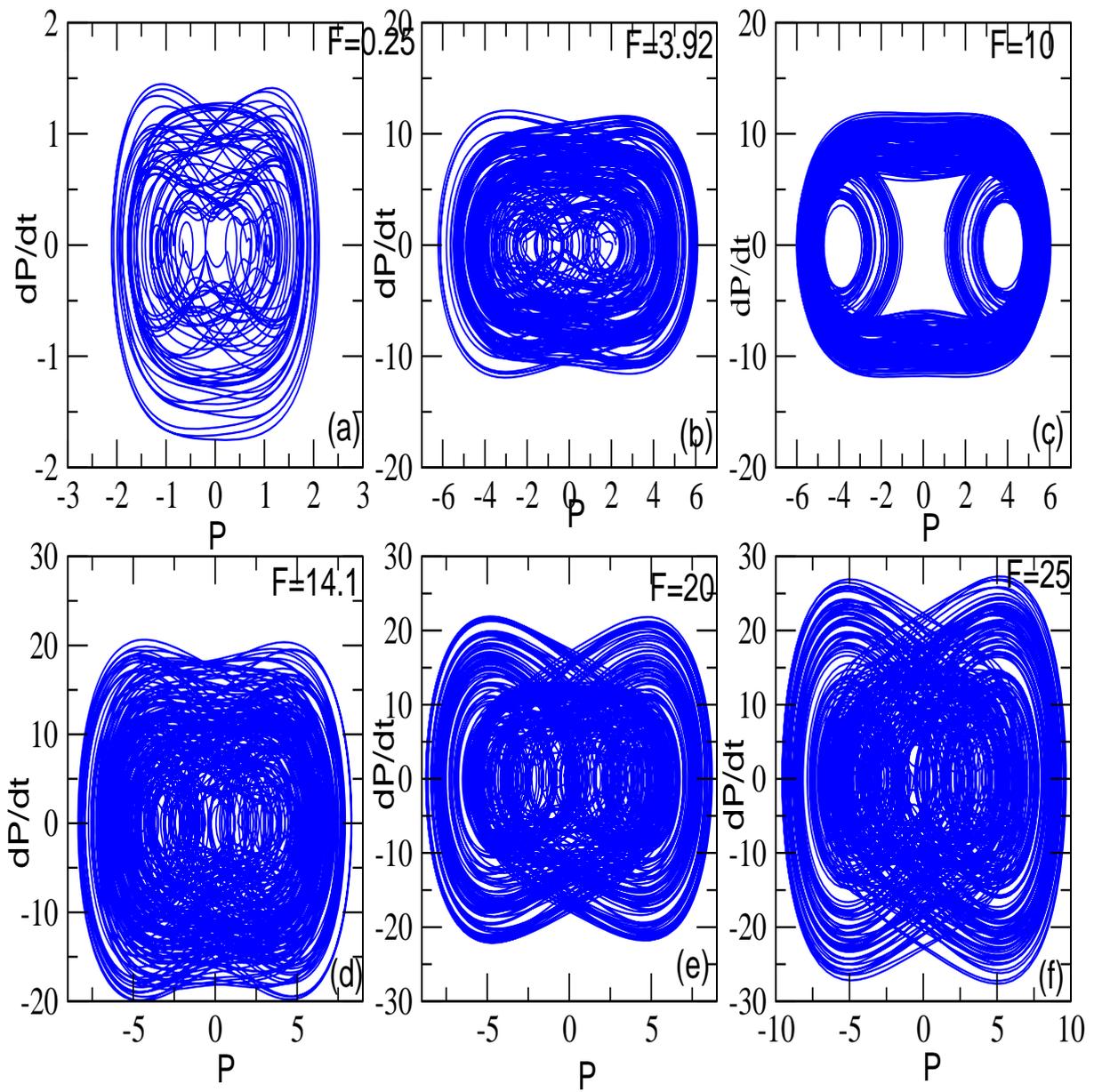

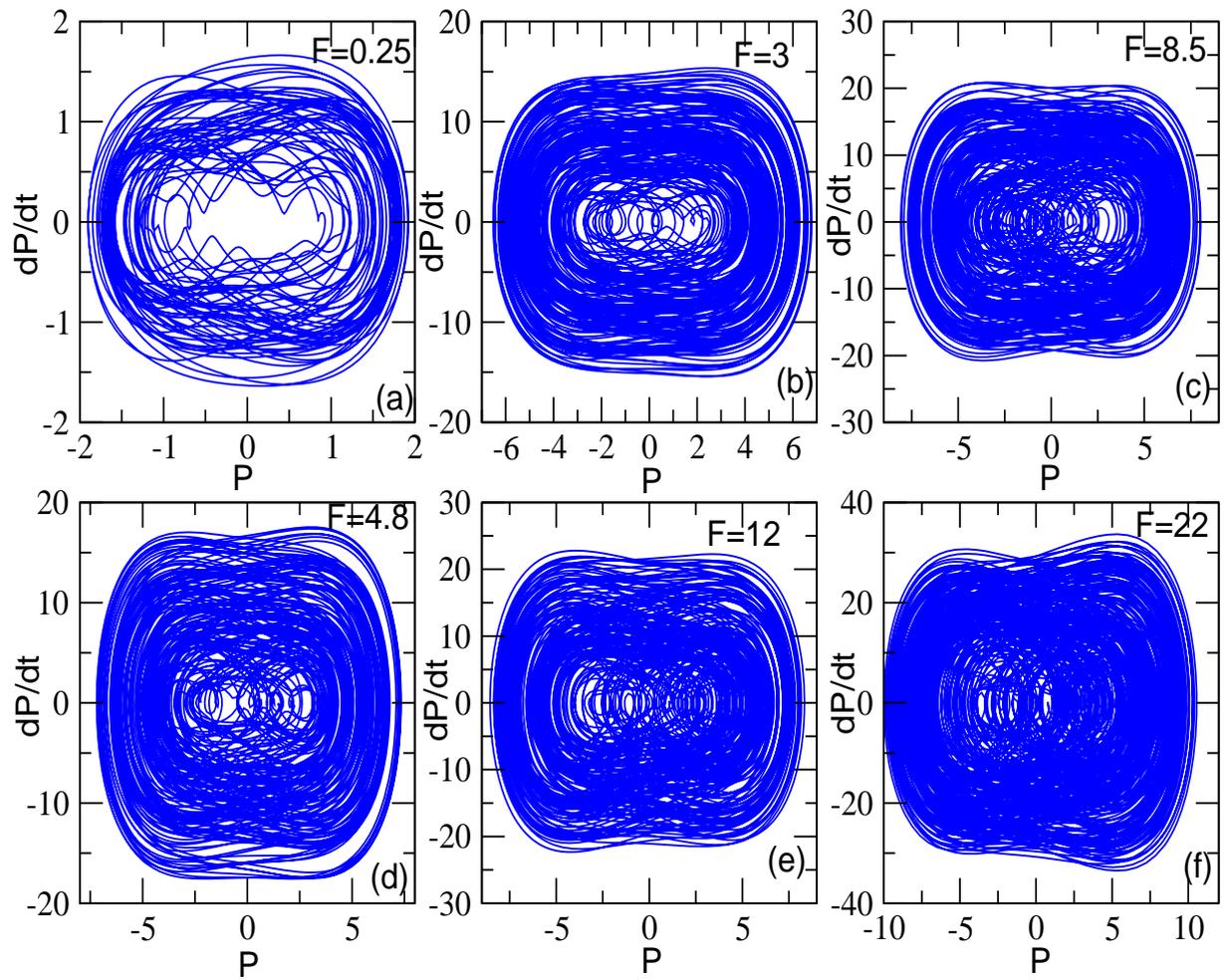

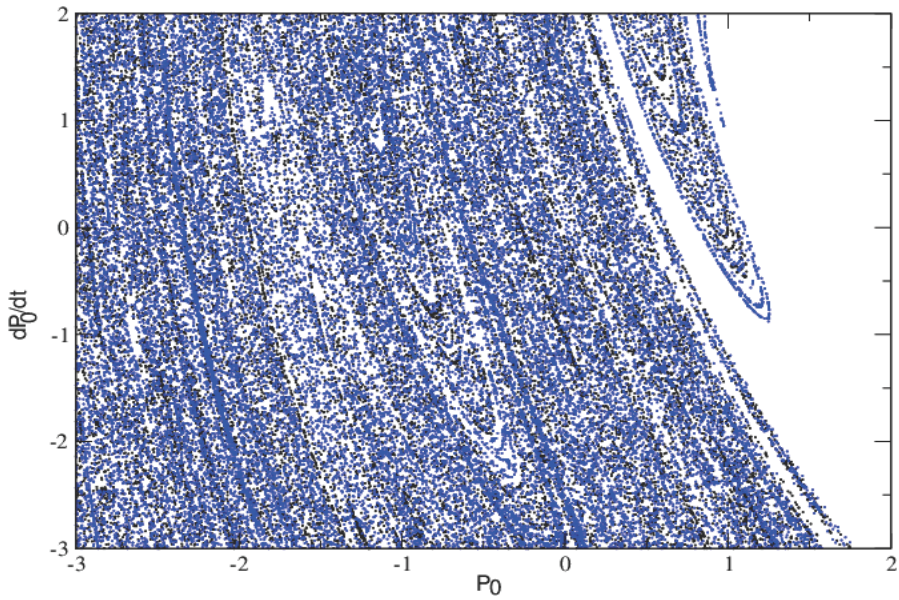

**Figure 11 :** Basin of chaoticity in the primary resonant state with F=16.5.

**(a)**

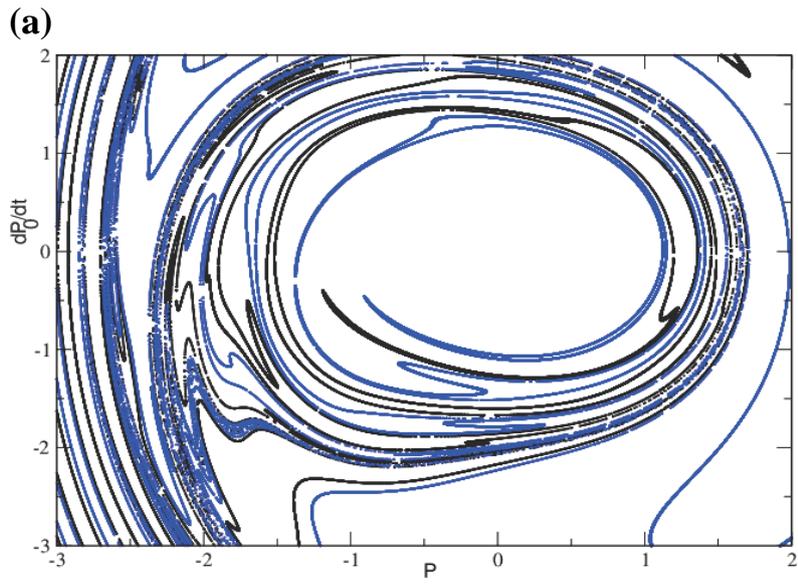

**(b)**

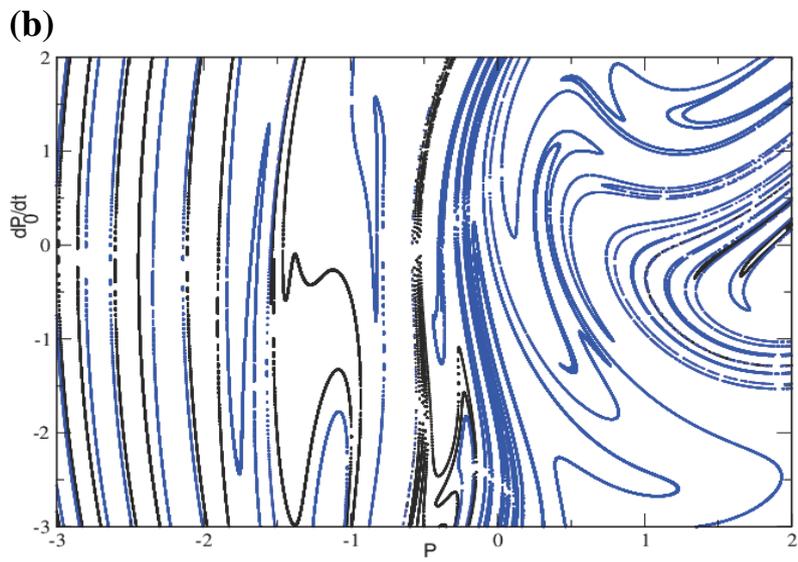

**(c)**

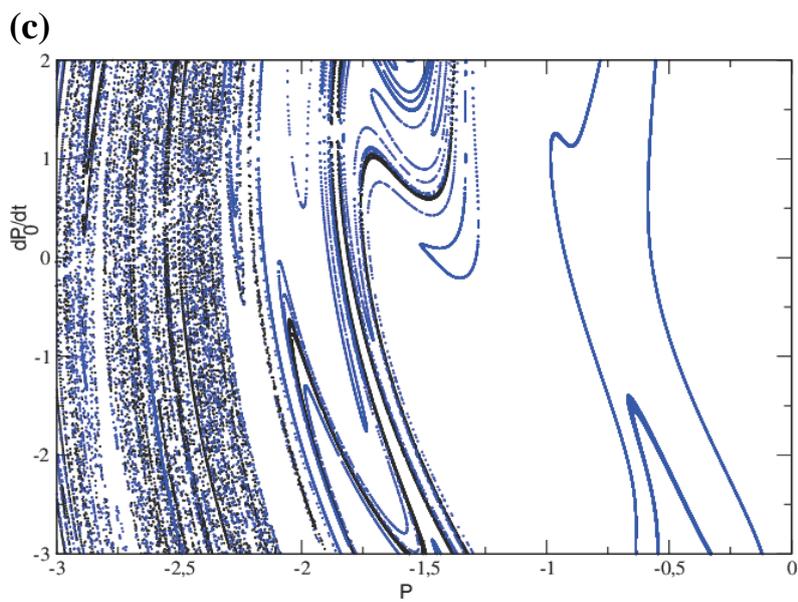

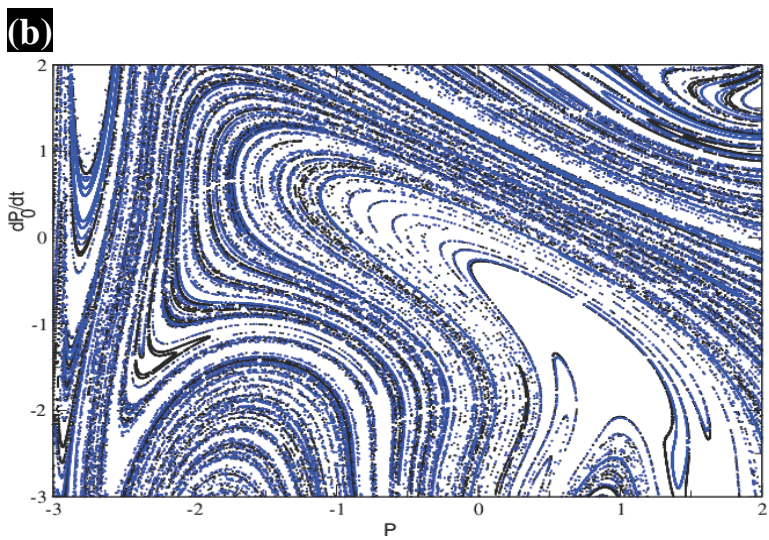

Figure12 : Basin of chaoticity in the secondary resonant state $w_1 + w_2 \simeq 2$ $(a) F = 0.25, (b) F = 3.92, (c) F = 10, (d) F = 14.1$.

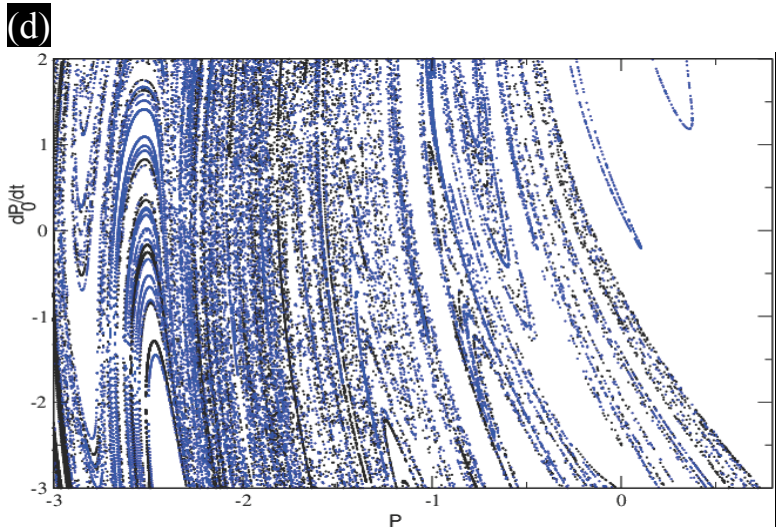

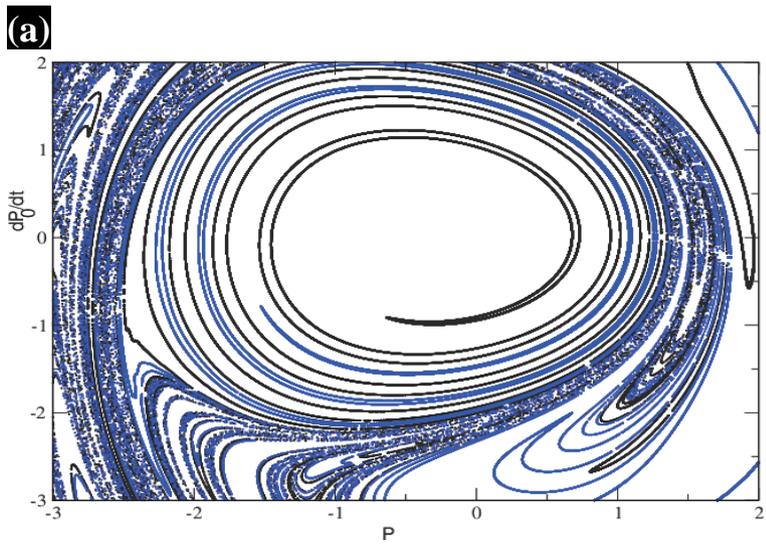

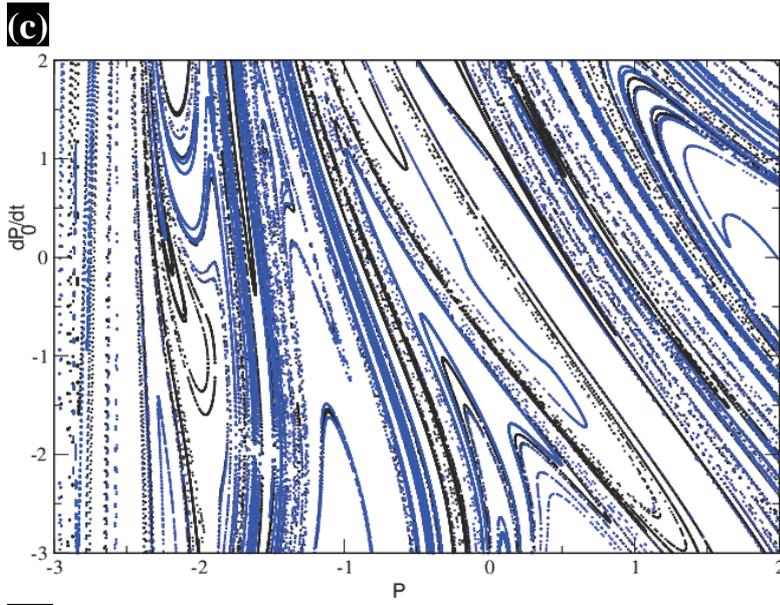
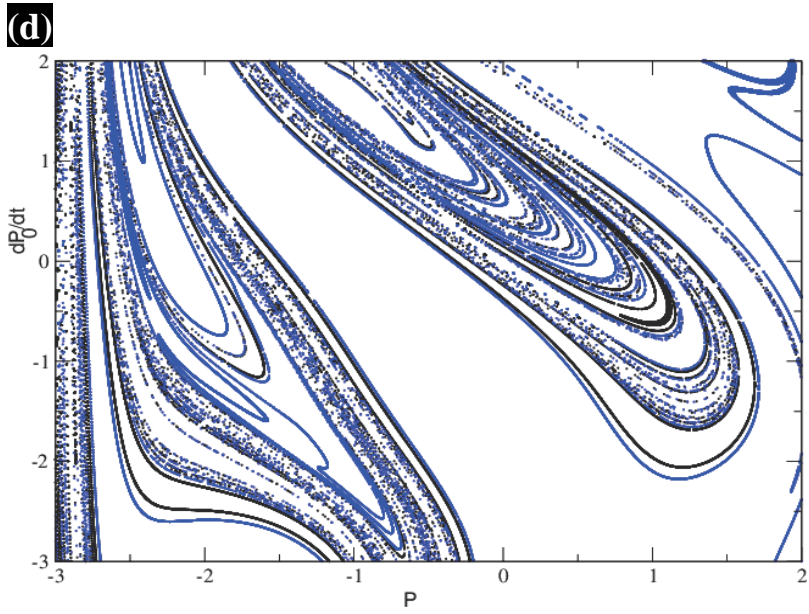

**Figure13 :** Basin of chaoticity in the secondary resonant state $2w_1 - w_2 \simeq 1$; $(a)F = 0.25, (b)F = 3, (c)F = 8.5, (d)F = 4.8$.

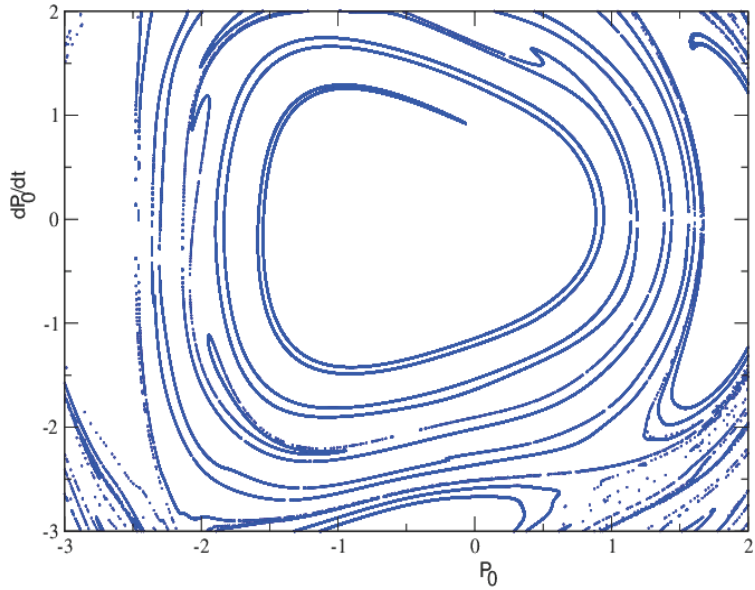

**Figure14 :** Basin of chaoticity in the first subharmonic resonant state with F=20.

## 4- Conclusion

In this paper, we have analysed the multiresonance and chaotic behavior of nonlinear dynamics of polarization oscillations when some materials when they are subjected to the action of an electromagnetic wave modeled by multifrequency forced Duffing equation. It is obtained by using multiple time scale method apart the cases of superharmonic ($3\omega_1 = 3\omega_2 = 1 + \sigma\varepsilon$) and *and subharmonic* ($\omega_1 = \omega_2 = 3 + \sigma\varepsilon$ treated in previous work [13], primary resonance and three cases for secondary resonances: $\omega_1+\omega_2 = 2$, $2\omega_2 - \omega_1 = 1$ *and* $2\omega_1 - \omega_2 = 1$. The jump and hysteresis phenomena are obtained for polarization oscillations. We noticed that for some of these phenomena the system has important singularities which can cause chaotic see catastrophic behavior for the material. This is confirm by bifurcation diagram and its corresponding Lyapunov exponent. It is found that the polarization oscillations is highly sensitive to the initial conditions through the basin of attraction which are obtained with the external forced amplitude predicted by different bifurcation diagrams.

**Acknowledgments**: Authors thank EDSM -UAC, IMSP- UAC, ICTP and the state of Benin for financial support. We also thank Laurent Hinvi for his fruitfull suggestions and Professor Paul Woafo for his suggestions and collaboration.